\documentclass{aa}
\usepackage{graphicx}
\def\approxgt{\mathrel{\hbox{\rlap{\lower.55ex \hbox {$\sim$}}
        \kern-.3em \raise.4ex \hbox{$>$}}}}
\def\approxlt{\mathrel{\hbox{\rlap{\lower.55ex \hbox {$\sim$}}
        \kern-.3em \raise.4ex \hbox{$<$}}}}

\begin{document}
   \title{X-ray obscuration and obscured AGN in the local universe}

   \author{Matteo Guainazzi
          \inst{1},
	  Giorgio Matt
	  \inst{2},
          Giuseppe Cesare Perola
          \inst{2}
          }

   \offprints{M.Guainazzi}

   \institute{$^1$XMM-Newton Science Operations Center, European Space Astronomy Center, ESA, Apartado
              50727, E-28080 Madrid, Spain \\
              \email{mguainaz@xmm.vilspa.esa.es} \\
	      $^2$Universit\`a degli Studi ``Roma Tre'', Via della Vasca Navale 84, I-00146, Roma, Italy
              }

   \date{Received 16 June 2005; accepted 04 August 2005}

   \abstract{In this paper we discuss the X-ray properties of
             49 local ($z<0.035$) Seyfert~2 galaxies with HST/WFC2 high-resolution optical coverage.
             It includes the results of 26 still unpublished {\it Chandra} and XMM-Newton observations,
             which yield 25 (22) new X-ray detections in the 0.5--2~keV (2--10~keV) energy band.
             Our sample covers a range in the 2--10 keV observed flux from $3 \times 10^{-11}$ to
             $6 \times 10^{-15}$~erg~cm$^{-2}$~s$^{-1}$. The percentage of the objects
             which are likely obscured by Compton-thick matter (column density, $N_H \ge \sigma_t^{-1}
             \simeq 1.6 \times 10^{24}$~cm$^{-2}$) is $\simeq$50\%, and reaches $\simeq$80\% 
             for $\log (F_{2-10}) < 12.3$. Hence, K$_{\alpha}$ fluorescent iron lines with large
             Equivalent Width ($EW > 0.6$~keV) are common in our sample (6 new detections at a
             confidence level  $\ge$2$\sigma$). They are explained as due to reflection off the illuminated side
	     of optically thick material.
	     We confirm 
             a correlation between the presence of a $\sim$100-pc scale nuclear dust in the WFC2 images
	     and Compton-thin obscuration.
             We interpret this correlation as due to the large covering fraction of gas associated with
             the dust lanes
             (following an idea originally proposed by Malkan et al. 1998, and Matt 2000). The X-ray spectra of
             highly obscured AGN invariably present a prominent soft excess emission above
	     the extrapolation of the hard X-ray component.
             This soft component can account for a very large fraction of the overall X-ray energy
             budget. As this component is generally unobscured - and therefore likely produced
             in extended gas structures - it may lead to a severe
             underestimation of the nuclear obscuration in $z \sim 1$ absorbed AGN,
             if standard X-ray colors are used to classify them. As a by-product of our study, we
             report the discovery of a soft X-ray, luminous
             ($\simeq$7$\times 10^{40}$~erg~s$^{-1}$) halo embedding the interacting
             galaxy pair Mkn~266.
            }

\authorrunning{Guainazzi et al.}

\titlerunning{X-ray obscured AGN in the local universe}

   \maketitle
%

\section{Introduction}

While it is widely accepted that ``type~2'' Active
Galactic Nuclei (AGN) are generally
obscured in X-rays by large column densities, $N_H$, the exact
location and distribution of the obscuring
gas are still debated. Traditional wisdom 
identifies the obscuring gas with matter responsible
for orientation-dependent effects ({\it e.g.}, the dusty
compact molecular ``torus'' of the AGN Unification
Scenarios; \cite{antonucci85,antonucci93}). However,
recent observational results challenge this
interpretation. Fast changes in the
X-ray column density advocate rather for 
matter closer to the nuclear supermassive black hole,
such as the Broad Line Region (BLR; \cite{elvis04,lamer03}).
``Narrow'' components  of
the K$_{\alpha}$ fluorescent iron
lines in unobscured AGN - when resolved by high-resolution
spectroscopic {\it Chandra} measurements - are sometimes
consistent with an origin in the BLR as well
(\cite{yaqoob01}, 2004; \cite{bianchi03b}), providing evidence
for the existence of obscuring matter on that
scale. However, both the low covering fraction of the BLR
and their optical thickness
seem too low to account for the large fraction of heavily
X-ray obscured AGN (e.g. \cite{risaliti99}). Other
pieces of evidence -
again mainly based on X-ray variability studies - suggest
a complex structure for the X-ray obscuring matter in many instances,  and in 
particular
rules out a single homogeneous, azimuthally-symmetric
obscuring structure (\cite{matt03a,guainazzi05a}).
Although an inhomogeneous ``patchy'' torus (\cite{pier92})
is still consistent with the data, and with the presence
of Compton-thick reflection in unobscured AGN
(\cite{bianchi04}), a possible scenario is
that X-ray obscuration occurs on various scales,
sometimes due to more compact matter close
to the black hole, sometimes due to
matter associated with the host galaxy rather then
with the nuclear environment. In
a few cases, ``multiple absorbers'' are directly
measured in X-rays (\cite{matt01}, 2003a), and their
lowest column densities are consistent with the optical
reddening due to host galaxy dust lanes,
once a standard Galactic gas-to-dust ratio
is assumed\footnote{On a possible
systematic difference in the gas-to-dust ratio
between our Galaxy and luminous nearby AGN,
see \cite{granato97}, and references therein}.

Another possible approach to elucidate the location and geometrical
distribution of X-ray obscuring matter in AGN is
a comparison between the distribution of X-ray column densities
and indicators of reddening and/or extinction at other wavelengths.
Prior to the launch of the
major X-ray observatories {\it Chandra} and XMM-Newton this
task was hampered by the limited number of objects, for which
X-ray spectroscopic measurements were available.

Guainazzi et al. (2001; GO1 hereafter) studied the correlation
between X-ray obscuration and the presence of nuclear dust
in the innermost $\sim$100~pc around the nucleus, as detected
in the high-resolution WFC2/HST imaging survey of nearby Seyfert galaxies
after Malkan et al. (1998; M98 hereafter). G01 discovered
a possible correlation: Compton-thin objects tend
to be preferentially located in galaxies with a larger dust content,
whereas the likelihood that a galaxy hosts a
Compton-thick nucleus is not affected by its
nuclear dust content.
However, the statistical significance of the
correlation in G01
was at the 1-$\sigma$ level. Since then, we have 
carried out a XMM-Newton observing program to verify
this tentative correlation on a larger
sample. The results of this program, coupled with the
analysis of still unpublished archival data of objects
belonging to the M98 sample, are the main subject of this
paper.

The structure of the paper is as follows: in Sect.~2 we introduce
the sample; in Sect.~3 we present
the reduction/analysis procedure. The
spectral results are described in Sect.~4, and
discussed in Sect.~5. Our main conclusions are
summarized in Sect.~6.

In this paper: the acceptance level for a spectral fit
corresponds to
a null hypothesis probability $\le 5\%$; errors on
the spectral fit parameters and thereof
derived quantities are at the 90\% confidence level
for 1 interesting parameter; all other uncertainties are
at the 1-$\sigma$ level; energies refer to the source
rest frame, unless otherwise specified.

\section{The sample}

The sample is made of all the Seyfert~2 galaxies in the Veron catalogue
(\cite{veron03}), which: a) have $z \le 0.035$; b) have been observed
in X-rays; c) have been included in the high-resolution
HST/WFC2 survey of M98. The plate scale of
WFC2 images (0$\arcsec$.046) corresponds
to a pixel size $\approxlt 32$~pc, and
the total image size (37$\arcsec$$\times$37$\arcsec$)
is $\approxlt 25$~kpc.  The total number of objects
fulfilling all three criteria is 49\footnote{We have
excluded from the sample UGC~4203 (the ``Phoenix Galaxy''), as
it belongs to the class of ``Changing-look''
Seyfert~2 galaxies (\cite{matt03b}), and its classification
with respect to the X-ray obscuration
is therefore ambiguous}, which represent
$\simeq$16\% of the total parent sample. 20 of
them were discussed by G01.
Although updated
measurements for the GO1 sample have been used,
whenever published in the literature, we {\it
did not} systematically reanalyze all GO1 sources, as they
are mostly well studied, X-ray bright AGN. Their
X-ray absorption classification - originally
derived from Bassani et al. (1999) - is
therefore robust.
The X-ray spectra of three more M98 objects 
have been published elsewhere:
ESO~509-IG066 (\cite{guainazzi05b}),
NGC~424 (\cite{matt03a}), NGC~7130 (\cite{levenson05}).
We present in Sect.~4 the results of
{\it Chandra}/XMM-Newton observations of the 
26 still unpublished objects belonging to the sample.

Following the classification in M98
we consider ``nuclear dusty'' those
galaxies, whose nuclear dust morphology falls
into the ``DC'' (``Dust
passing close or crossing the Center'') and ``DI''
(``Dust Irregular'')\footnote{{\it i.e.}, ``not
associated with any spiral arm pattern'', M98}
categories. We have verified that the adopted
classification is consistent with
that reported by other authors, which discuss
the same nuclear
images, or images with
comparable spatial resolution
(\cite{ferruit00,quillen99b,martini03}).
The mean redshifts for ``dusty'' and ``non-dusty''
objects are consistent: $\langle z \rangle = 0.016 \pm 0.009$
for the former,
$\langle z \rangle = 0.017 \pm 0.008$ for the latter,
respectively. The overwhelming majority of the
galaxies in our sample (86\%) are spirals.
The remaining are almost equally shared between
ellipticals (5\%) and irregulars (9\%).

The list of objects, for which new X-ray data are presented
in this paper, is reported in Tab.~\ref{tab1}, with redshifts,
$H_{\alpha}/H_{\beta}$ flux ratios, O[{\sc iii}] fluxes and
some information on the {\it Chandra}/XMM-Newton observation.
\begin{table*}
\caption{Properties of the Seyfert~2 galaxies discussed in this paper. The ``Date''
indicates the observation date. $T_{exp}$ is the net
exposure time in the pn (ACIS) XMM-Newton ({\it Chandra}) camera. The ``Threshold'' was
applied on the XMM-Newton EPIC E$> 10$~keV, field-of-view, single-event
light curves employed to remove intervals of high particle background. ``Radius'' refers
to the size of the source extraction region in MOS/pn (XMM-Newton
observations) and ACIS-S ({\it Chandra} observations), respectively.}
\begin{tabular}{lccccccccc} \hline \hline
Source & $z$ & $N_{H,Gal}$ & H$_{\alpha}$/H$_{\beta}$ & F$_{O[III]}$$^a$ & Ref.$^b$ & Date & $T_{exp}$ & Threshold & Radius \\
& & ($10^{20}$~cm$^{-2}$) & & & & (DD-MM-YY) & (ks) & (s$^{-1}$) & (arcsecs) \\ \hline
\multicolumn{10}{l}{Seyfert~2 exhibiting nuclear dust} \\
Mkn266 & 0.0281 & 1.7 & 5.89 & 0.32 & 8 & 02-11-01 & 19.7 & ... & 2.0 \\
Mkn612 & 0.0203 & 5.0 & 6.61  & 1.80 & 8 & 19-01-04 & 9.1 & 0.35/1.0 & 40/25 \\
NGC4156 & 0.0225 & 2.0  & 33.33 & ... & 5 & 22-12-00 & 51.2 & .../... & 40/40 \\
NGC4968 & 0.0099 & 9.2 & 14.9 & 1.00 & 8 & 05-01-01 & 4.3 & .../... & 40/40 \\
NGC7212 & 0.0266 & 5.5  & 5.01 & 7.08 & 6 & 20-05-04 & 10.4 & .../1.0 & 30/60 \\
NGC7465 & 0.0066 & 6.0  & 5.34 & ... & 1 & 29-07-03 & 5.0 & ... & 6.0 \\
UGC2456 & 0.0120 & 12.1  & 8.51 & 2.40 & 8 & 20-02-05 & 11.7 & 0.35/2.0 & 35/35 \\
UGC4229 & 0.0232 & 5.2  & 18.20 & 0.40 & 6 & 05-05-03 & 4.8 & 0.35/1.0 & 20/40 \\
UGC6527 & 0.0274 & 1.1  & 6.57 & 5.40 & 6 & 04-11-04 & 7.8 & 0.35/1.0 & 50/40 \\
UGC987 & 0.0155 & 5.7  & 5.38 & 0.30 & 2 & 23-01-04 & 14.6 &  0.35/1.0 & 50/40 \\ \hline
\multicolumn{10}{l}{Seyfert~2 {\it not} exhibiting nuclear dust} \\
ESO269-G012 & 0.0165 & 13.8 & ... & 0.50 & 7 & 06-01-04 & 17.6 & ... & 6.8 \\
{\footnotesize 2MASSJ01500266-0725482} & 0.0177 & 2.3 & 7.62 & 0.53 & 4 & 21-01-04 & 9.0 & 0.35/1.0 & 45/45 \\
IC~3639 & 0.0109 & 5.1 & 6.14 & 3.14 & 4 & 07-03-04 & 8.7 & ... & 5.2 \\
IC4995 & 0.0161 & 4.2 &  5.37 & 2.81 & 8 & 25-09-04 & 6.7 & 0.35/1.0 & 20/30 \\
IRAS15480-0344 & 0.0303 & 9.4  & 10.19 & 1.38 & 4 & 31-07-03 & 2.5 & 0.35/2.0 & 20/25 \\
NGC1410 & 0.0253 & 8.2  & 15.0 & 0.36 & 8 & 22-01-04 & 7.8 & 0.35/2.0 & 25/40 \\
Mkn1 & 0.0154 & 5.5  & 3.32 & 3.47 & 4 & 09-01-04 & 9.1 & .../2.0 & 45/30 \\
NGC17 & 0.0198 & 2.6  & 43.75 & 0.11 & 8 & 22-12-02 & 10.6 & 0.35/1.0 & 40/40 \\
UGC1214 & 0.0172 & 3.0  & 4.17 & 15.85 & 8 & 15-01-04 & 9.1 & .../... & 32/32 \\
UGC2608 & 0.0233 & 14.1  & 6.31 & 2.19 & 6 & 28-01-02 & 1.1 & 1.0/2.0 & 25/25 \\
NGC5283 & 0.0104 & 1.8  & 3.80 & 2.69 & 8 & 24-11-03 & 8.9 & ...  & 2.9 \\
NGC5427 & 0.0087 & 2.2 & ... & ... & ... & 26-03-04 & 8.8 & ... & ...$^c$ \\
NGC5953 & 0.0066 & 3.3  & 22.0 & 8.60 & 5,9 & 29-12-02 & 9.9 & ... & 2.5 \\
UM625 & 0.0250 & 4.0  & 9.17 & 2.10 & 3 & 02-07-04 & 7.4 & .../1.0 & 30/30 \\
NGC3982 & 0.0037 & 1.2  & 4.00 & 1.51 & 8 & 03-01-04 & 9.2 & ... & 3.7 \\
NGC591 & 0.0152 & 4.8 & 6.02 & 2.30 & 6 & 12-01-04 & 8.8 & 0.35/1.0 & 25/40 \\ \hline
\hline \hline
\end{tabular}

\noindent
$^a$in units of $10^{-13}$~erg~cm$^{-2}$~s$^{-1}$

\noindent
$^b$1.- Osterbrock et al. (1987); 2.- Dahari et al. (1988); 3.- Terlevich et al. (1991); 4.- de Grijp et al. (1992); 5.- Kennicutt et al. (1992);
6.- Mulchaey et al. (1994); 7.- Mulchaey et al. (1996); 8.- Polletta et al. (1996); 9.- Risaliti et al. (1999)

\noindent
$^c$not detected

\label{tab1}
\end{table*}

\section{Observations and data reduction}

The XMM-Newton data 
were reduced with SAS v6.1
(\cite{gabriel03}), using the most updated calibration
files available at the moment the data reduction
was performed. In this paper, only data from the
EPIC cameras (MOS; Turner et al. 2001; pn,
Str\"uder et al. 2001) will be discussed.
Event lists from the two MOS cameras were
merged before accumulation of any scientific
products. Single to double (quadruple)
events were used to accumulate pn (MOS) spectra.
High-background particle flares were removed, by
applying fixed thresholds on the single-event,
$E>10$~keV, $\Delta t = 10$~s light curves.
These thresholds, as well as the radius of the
source circular extraction regions,
were optimized to maximize the signal-to-noise
ratio, and are listed in Tab.~\ref{tab1}.
The background  
was extracted from annuli around the source
for the MOS, and from
circular regions in the same chip
for the pn, at the same height in detector coordinates as the
source location.
Spectra were binned
in order to oversample the intrinsic
instrumental energy resolution by a factor
$\sim$3.
If the resulting spectral bins had less
than 25 background-subtracted counts, the
spectra were further rebinned to reach
this limit, which ensures the
applicability of the $\chi^2$ statistics
to evaluate the quality of the
spectral fitting. pn (MOS) spectra were
fitted in the 0.35--15~keV (0.5--10~keV)
spectral range.

The {\it Chandra} data were reduced with
{\sc Ciao 3.1}, starting from the linearized
``level 2'' event lists, using
{\sc Caldb v.2.28}. The ACIS-S spectra
have been rebinned according to the same
criteria employed for the EPIC ones. Fits
were performed in the 0.1--11~keV energy
band.

Out of the 26 objects in Tab.~\ref{tab1},
25 (22) were detected at a level larger then 3$\sigma$
in the 0.5--2~keV (2--10~keV) energy band.

\section{Spectral results}

\begin{figure*}
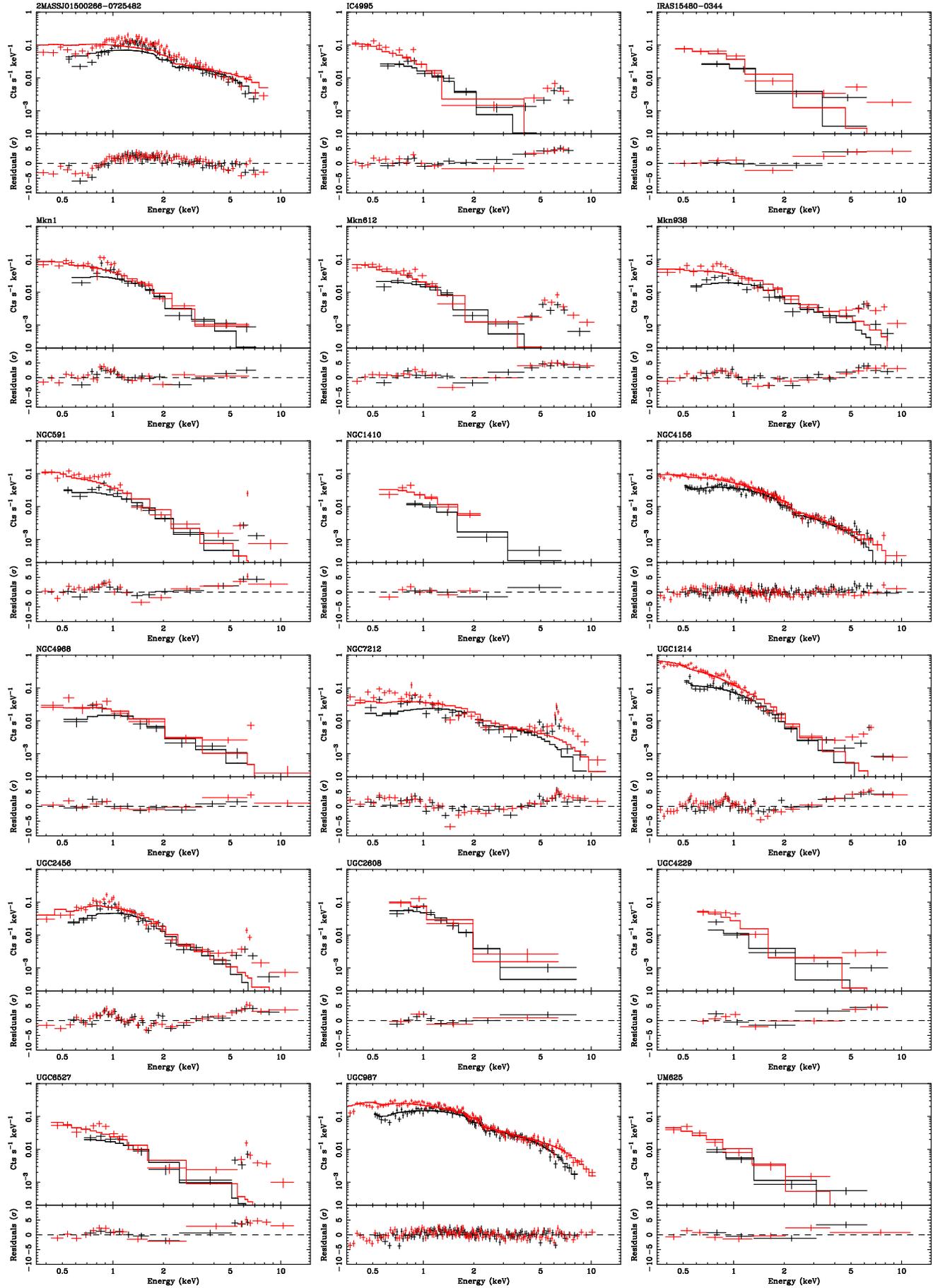

\hbox{
 \includegraphics[width=4cm, angle=-90]{fig1a.ps}
 \includegraphics[width=4cm, angle=-90]{fig1r.ps}
 \includegraphics[width=4cm, angle=-90]{fig1f.ps}
}
\hbox{
 \includegraphics[width=4cm, angle=-90]{fig1g.ps}
 \includegraphics[width=4cm, angle=-90]{fig1b.ps}
 \includegraphics[width=4cm, angle=-90]{fig1c.ps}
}
\hbox{
 \includegraphics[width=4cm, angle=-90]{fig1w.ps}
 \includegraphics[width=4cm, angle=-90]{fig1p.ps}
 \includegraphics[width=4cm, angle=-90]{fig1e.ps}
}
\hbox{
 \includegraphics[width=4cm, angle=-90]{fig1z.ps}
 \includegraphics[width=4cm, angle=-90]{fig1q.ps}
 \includegraphics[width=4cm, angle=-90]{fig1j.ps}
}
\hbox{
 \includegraphics[width=4cm, angle=-90]{fig1d.ps}
 \includegraphics[width=4cm, angle=-90]{fig1m.ps}
 \includegraphics[width=4cm, angle=-90]{fig1k.ps}
}
\hbox{
 \includegraphics[width=4cm, angle=-90]{fig1t.ps}
 \includegraphics[width=4cm, angle=-90]{fig1u.ps}
 \includegraphics[width=4cm, angle=-90]{fig1s.ps}
}
\caption{Spectra ({\it upper panels}) and
residuals against a power-law continuum ({\it lower panels})
for the XMM-Newton/EPIC observations of our sample.}
\label{fig1}
\end{figure*}

\begin{figure*}
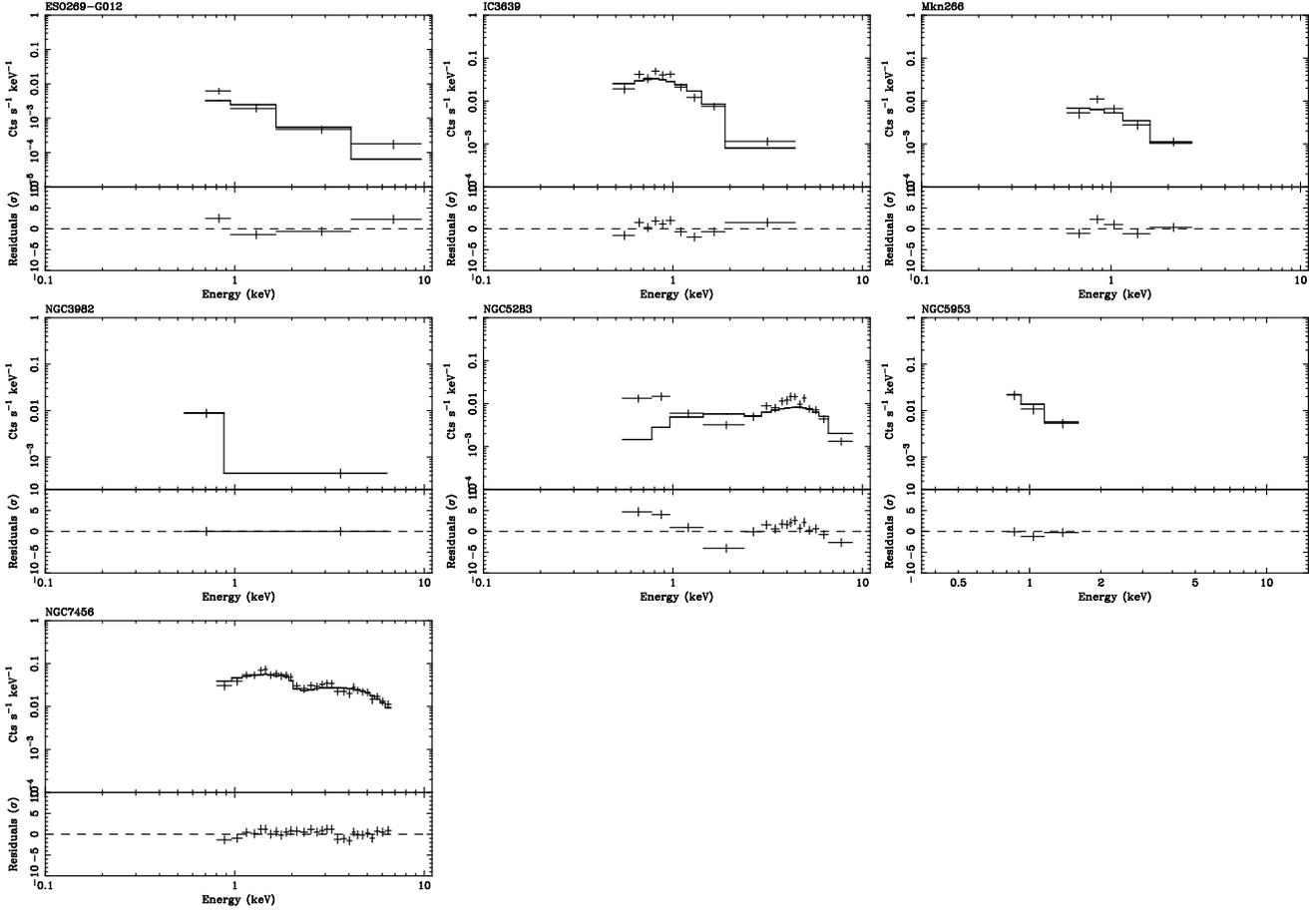

\hbox{
 \includegraphics[width=4cm, angle=-90]{fig3x.ps}
 \includegraphics[width=4cm, angle=-90]{fig3aa.ps}
 \includegraphics[width=4cm, angle=-90]{fig3a.ps}
}
\hbox{
 \includegraphics[width=4cm, angle=-90]{fig3v.ps}
 \includegraphics[width=4cm, angle=-90]{fig3e.ps}
 \includegraphics[width=4cm, angle=-90]{fig3b.ps}
}
\hbox{
 \includegraphics[width=4cm, angle=-90]{fig3c.ps}
}
\caption{Spectra ({\it upper panels}) and
residuals against a power-law continuum ({\it lower panels})
for the {\it Chandra}/ACIS observations of our sample.}
\label{fig5}
\end{figure*}

Previous
studies show that X-ray spectra of obscured AGN
are complex (\cite{turner97, risaliti02, guainazzi05a}). This holds for the
objects discussed in this paper as well. In Fig.~\ref{fig1}
and~\ref{fig5} we
show the sample source spectra, and their residuals when a
photoelectrically absorbed power-law model is
applied. This simple model leaves significant residuals in
almost all
objects. As already discussed by, {\it e.g.}, Guainazzi
et al. (2005a), at least two continua components in the 0.5--10~keV energy
band are required.
Fluorescent emission lines from iron and other elements,
often covering a wide range of ionization stages, are often present as well,
as witnessed by local excesses in the
residuals of Fig.~\ref{fig1} and~\ref{fig5} at
$\simeq$0.6, 0.9, and 6.4~keV.

Hence, we have defined a set of
physically-motivated ``baseline'' models to fit the data.
For Compton-thin Seyfert~2s the baseline model
is expressed by the following general formula:
$$
F_{thin}(E) = M + f_s A E^{- \Gamma} + e^{-\sigma(E) N_H} A E^{- \Gamma} + \sum_i G_i(E)
$$
where $\sigma(E)$ is the photoelectric cross-section,
with solar abundances according to
Andres \& Grevesse (1989); $A$ is a normalization
factor; $f_s$ (``scattering fraction'') is
constrained: $0 \le f_s \le 1$; $M$ is the
thermal emission from a collisionally ionized plasma
with temperature $T$
(we used the {\tt mekal} implementation in
{\sc Xspec}; \cite{mewe85}); $G_i(E)$ are
Gaussian profiles, corresponding to fluorescence
emission lines of high-Z elements from O to
Fe, the last being by far the most important and
common with its 6.4~keV neutral fluorescent line.
For ``Compton-thick'' objects, the baseline model
is slightly different:
$$
F_{thick}(E) = M + B E^{- \Gamma} + R(E, \Gamma) + \sum_i G_i(E)
$$
where $B$ is a normalization constant,
and $R(E)$ is the ``bare'' Compton-reflection of
an otherwise invisible nuclear continuum
(we used the implementation {\tt pexrav} in {\sc
Xspec}, \cite{magdziarz95}). In this component,
we have assumed again solar abundances,
and a face-on slab. No high-energy
cut-off for the intrinsic power-law continuum
emission has been imposed, as it normally lays
beyond the sensitive bandpass of the XMM-Newton/EPIC
or {\it Chandra}/ACIS cameras (\cite{perola02}).
In principle, the Compton reflection component should have 
been added also to the baseline model for Compton-thin objects. 
However, its contribution in the pn energy
bandpass is expected in these cases to be negligible, and
was therefore not included for simplicity.

Observed fluxes in the 0.5--2.0~keV and
2--10~keV energy bands, and absorption-corrected luminosity for
the thermal and the AGN power-law components are shown in
Tab.~\ref{tab3}.
\begin{table*}
\caption{X-ray observed fluxes, $T$ values, and absorption-corrected luminosities in the 0.5--10~keV energy range}
\begin{tabular}{lcccccc} \hline \hline
Source & \multicolumn{3}{c}{Observed fluxes} & & \multicolumn{2}{c}{Luminosities} \\
& 0.5--2~keV$^a$ & 2--10~keV$^a$ & 5--10~keV$^b$ & $T$ & Thermal$^c$ & AGN$^d$ \\ \hline
2MASSJ01500266-0725482 & 0.30 & 0.82 & 3.7 &1.00 & $0.9 \pm^{1.7}_{0.6}$ & $1.20 \pm^{0.06}_{0.07}$ \\
ESO209-G012 & 0.011 & 0.06 & 4.2 & 1.2$^e$ & $1.0 \pm 0.3$ & $0.13 \pm^{0.10}_{0.07}$ \\
IC3639 & 0.10 & 0.08 & 0.39 & 0.03 & $2.4 \pm^{0.9}_{1.4}$ & ... \\
IC4995 & 0.06 & 0.29 & 2.5 & 0.18 & $4 \pm 2$ & ... \\
IRAS15480-0344 & 0.09 & 0.37 & 2.8 & 0.07 & $27 \pm^{233}_{15}$ & ... \\
Mkn1 & 0.10 & 0.13 & 1.0 & 0.28 & ... & $1.29 \pm^{5.50}_{1.28}$ \\
Mkn266 & 0.03 & $<$0.007 & $<$0.03 & $<$0.03 &  $3.6 \pm^{1.1}_{1.0}$ & ... \\
Mkn612 & 0.05 & 0.36 & 3.2 & 0.20 & $2.7 \pm^{0.6}_{0.5}$ & $7.3 \pm^{1.3}_{1.1}$ \\
NGC17 & 0.07 & 0.33 & 2.0 & 0.01 & $2.0 \pm 0.5$ & $2.16 \pm 0.06$ \\
NGC591 & 0.09 & 0.20 & 1.6 & 0.11 & $3.4 \pm^{1.9}_{0.9}$ & ... \\
NGC1410 & 0.04 & 0.04 & 0.2 & 0.01 & $2.3 \pm^{1.8}_{0.9}$ & $0.10 \pm ^{0.03}_{0.02}$ \\
NGC3982 & 0.018 & $<$0.05 & $<$0.2 & $<0.14$ & $0.057 \pm 0.0014$ & ... \\
NGC4156 & 0.13 & 0.18 & 0.8 & ... & ... & $0.414 \pm 0.015$ \\
NGC4968 & 0.04 & 0.15 & 1.2 & 0.04 & ... & ... \\
NGC5283 & 0.04 & 1.46 & 10.3 & 2.71 & ... & $1.7 \pm^{0.8}_{0.6}$ \\
NGC5427 & $<$0.002 & $<$0.11 & $<$0.11 & ... & ... & ... \\
NGC5953 & 0.06 & $<$0.006 & $<$0.02 & $<0.01$ & ... & ... \\
NGC7212 & 0.09 & 0.69 & 5.6 & 0.22 & $8 \pm^4_2$ & ... \\
NGC7465 & 0.24 & 4.15 & 29.0 & ... & ... & $1.0 \pm^{0.8}_{0.3}$ \\
UGC1214 & 0.32 & 0.28 & 1.7 & 0.07 & $16 \pm^6_3$ & ... \\
UGC2456 & 0.14 & 0.36 & 2.8 & 0.07 & ... & ... \\
UGC2608 & 0.13 & 0.14 & 0.8 & 0.03 & $30 \pm^9_{11}$ & ... \\
UGC4229 & 0.05 & 0.22 & 1.7 & 0.03 & $4.1 \pm^{1.2}_{1.1}$ & ... \\
UGC6527 & 0.05 & 0.43 & 4.0 & 0.08 & $3.2 \pm^{1.1}_{1.0}$ & $15 \pm^{14}_7$ \\
UGC987 & 0.48 & 1.21 & 6.1 & 7.24 & ... & $1.09 \pm 0.05$ \\
UM625 & 0.03 & 0.02 & 0.11 & 0.004 & $2.3 \pm^{2.5}_{1.0}$ & ... \\
\hline \hline
\end{tabular}

\noindent
$^a$in units of $10^{-12}$~erg~cm$^{-2}$~s$^{-1}$

\noindent
$^b$in units of $10^{-13}$~erg~cm$^{-2}$~s$^{-1}$

\noindent
$^c$in units of $10^{40}$~erg~s$^{-1}$

\noindent
$^d$in units of $10^{42}$~erg~s$^{-1}$

\noindent
$^e$not corrected for Balmer decrement (measurement not available in the literature)

\label{tab3}
\end{table*}

The spectra of the large majority of
the sources discussed in this paper are comparatively
weak, with an overall
number of independent spectral channels often lower than 50.
In most cases, not all the components of the baseline
model are required. Moreover, in many objects
it is not possible
to discriminate whether the object is Compton-thick
or -thin from the pure statistical point of view. In
order to proper classify these objects, we need to
complement the statistical results of the fits with
the evaluation of
two physical observables with a high diagnostic power 
in X-ray obscured AGN: the Equivalent Width ($EW$)
of the Fe K$_{\alpha}$ fluorescent line, and the
ratio $T$ between the 2--10~keV flux and the Balmer-Decrement
corrected O[{\sc iii}] flux.

\begin{figure*}
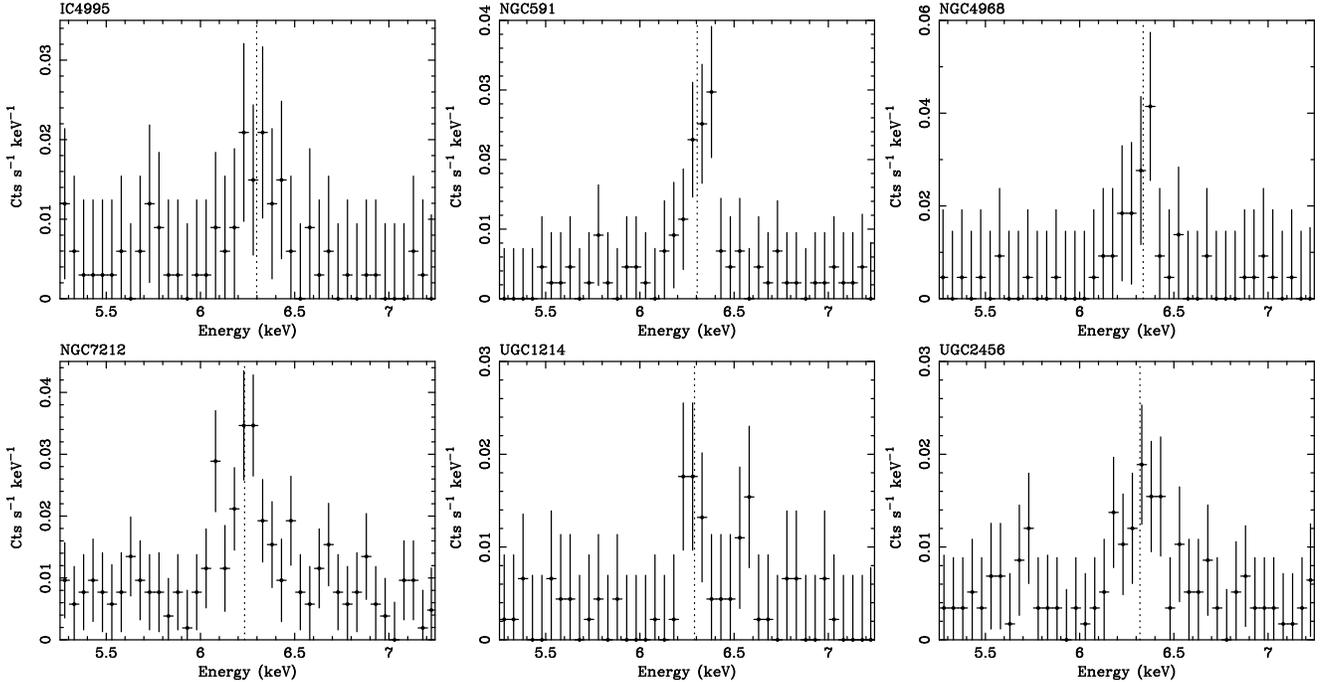

\hbox{
 \includegraphics[width=4.5cm, angle=-90]{fig2r.ps}
 \includegraphics[width=4.5cm, angle=-90]{fig2w.ps}
 \includegraphics[width=4.5cm, angle=-90]{fig2z.ps}
}
\hbox{
 \includegraphics[width=4.5cm, angle=-90]{fig2q.ps}
 \includegraphics[width=4.5cm, angle=-90]{fig2j.ps}
 \includegraphics[width=4.5cm, angle=-90]{fig2l.ps}
}
\caption{Zoom of the spectra in the 5.25--7.25~keV
(observer's frame) for the sources, where the
K$_{\alpha}$ iron line is detected at a confidence level
$\ge$2$\sigma$. The spectra are linearly binned
with a constant $\delta E = 50$~eV. Each channel
corresponds therefore approximately to one-third of
a resolution element.}
\label{fig2}
\end{figure*}

The binning requirement imposed by the application of the
$\chi^2$ test in the global fits may hamper the 
detectability of even large EW emission lines in low-statistics spectra with almost
no continuum besides the line. 
Before applying the baseline models
to the broadband X-ray spectra,
we have therefore fit the {\it unbinned} spectra in the 5.25-7.25~keV
energy range (where the $K_{\alpha}$ fluorescent iron line complex
is expected to lay for all objects in our sample), using
the Cash statistics (\cite{cash76}; ``local fits'' hereinafter).
Power-laws were
used to independently model the continuum and the background, as
the Cash method cannot be applied on 
spectra, which do not follow the Poisson
statistics. Their
parameters were then allowed
to vary in the local fits within their confidence intervals.
The properties of the iron K$_{\alpha}$ lines
derived from the local fits are summarized in Tab.~\ref{tab4}.
\begin{table*}
\caption{Fe K$_{\alpha}$ properties}
\begin{tabular}{lccccc} \hline \hline
& \multicolumn{2}{c}{Local fits} & \multicolumn{3}{c}{Global fits} \\
Source & $E_l$ & $EW$ & $E_l$ & $EW$ & $I_l$$^{\ddag}$ \\
& (keV) & (keV) & (keV) & (keV) & \\ \hline
2MASSJ01500266-0725482 & $6.4$$^{\dag}$ & $< 200$ & $6.56 \pm^{0.19}_{0.14}$ & $350 \pm^{210}_{290}$ & 0.23 \\
ESO269-G012 & $6.4$$^{\dag}$ & $<700$ & $6.4$$^{\dag}$ & $<3000$ & $<0.3$ \\
IC3639 & $6.39 \pm 0.07$ & $1500 \pm 1100$ & $6.4$$^{\dag}$ & $<$525000 & $<$174 \\
IC4995 & $6.42 \pm^{0.03}_{0.05}$ & $1700 \pm 700$ & $6.40 \pm^{0.03}_{0.05}$ & $4300 \pm^{1400}_{1100}$ & 1.1 \\
IRAS15480-0344 & $6.41 \pm^{0.03}_{0.04}$ & $17000 \pm^{8000}_{10000}$ & $6.4$$^{\dag}$ & $<2400$ & $<1.5$ \\
Mkn1 & $6.4$$^{\dag}$ & $<800$ & $6.4$$^{\dag}$ & $<2000$ & $<0.5$ \\
Mkn266 & $6.38 \pm^{0.07}_{0.06}$ & $6000 \pm^{2000}_{4000}$ & 6.4$^{\dag}$ & $<100000$ & $<50$ \\
Mkn612 & $6.46 \pm^{0.02}_{0.09}$ & $280 \pm^{240}_{180}$ & $6.40 \pm 0.18$ & $200 \pm^{360}_{160}$ & 0.5 \\
NGC17 & $6.4$$^{\dag}$ & $<1900$ & $6.4$$^{\dag}$ & $<322$ & $<0.7$ \\
NGC591 & $6.41 \pm^{0.03}_{0.02}$ & $2200 \pm^{700}_{600}$ & $6.47 \pm 0.03$ & $5800 \pm^{1400}_{2300}$ & 0.9 \\
NGC1410 & $6.420 \pm 0.010$ & $800 \pm^{700}_{600}$ & $6.4$$^{\dag}$ & $<11000$ & $<0.5$ \\
NGC3982 & $6.4$$^{\dag}$ & $8000 \pm 5000$ & $6.4$$^{\dag}$ & $<41000$ & $<3.0$ \\
NGC4156 & $6.4$$^{\dag}$ & $<300$ & $6.4$$^{\dag}$ & $<500$ & $<0.10$ \\
NGC4968 & $6.40 \pm^{0.02}_{0.03}$ & $3000 \pm^{1200}_{1000}$ & $6.41 \pm 0.02$ & $5300 \pm^{1700}_{2200}$ & 1.8 \\
NGC5283 & $6.4$$^{\dag}$ & $<170$ & $6.4$$^{\dag}$ & $<220$ & $<0.7$ \\
NGC5953 & ... & ... & ... & ... & ... \\
NGC7212 & $6.41 \pm 0.03$ & $900 \pm^{200}_{300}$ & $6.41 \pm^{0.03}_{0.02}$ & $1100 \pm 200$ & 0.9 \\
NGC7465 & $6.4$$^{\dag}$ & $<300$ & $6.4$$^{\dag}$ & $<400$ & $<3$ \\
UGC1214 & $6.41 \pm 0.05$ & $1300 \pm 500$ & $6.41 \pm 0.05$ & $2700 \pm 1000$ & 0.6 \\
& $6.66 \pm 0.05$ & $700 \pm^{600}_{300}$ &$6.92 \pm 0.09$ & $15000 \pm^{5000}_{8000}$ & 0.4 \\
UGC2456 & $6.41 \pm^{0.04}$ & $1000 \pm^{300}_{400}$ & $6.42 \pm^{0.04}_{0.05}$ & $320 \pm 90$ & 0.6 \\
UGC2608 & $6.4$$^{\dag}$ & $<1300$ & $6.4$$^{\dag}$ & $<8000$ & $<1.2$ \\
UGC4229 & $6.4$$^{\dag}$ & $<4000$ & $6.4$$^{\dag}$ & $<1600$ & $<0.7$\\
UGC6527 & $6.37 \pm^{0.08}_{0.05}$ & $350 \pm^{230}_{190}$ & $6.44 \pm 0.08$ & $100 \pm 50$ & 0.3 \\
UGC987 & $6.45 \pm 0.07$ & $180 \pm^{110}_{100}$ & $6.48 \pm^{0.24}_{0.07}$ & $120 \pm 80$ & 0.18 \\
UM625 & $6.4$$^{\dag}$ & $<1100$ &  $6.4$$^{\dag}$ & $<6000$ & $<0.2$ \\
\hline \hline
\end{tabular}

\noindent
$^{\ddag}$in units of $10^{-5}$~ph~cm$^{-2}$~s$^{-1}$

\noindent
$^{\dag}$fixed

\label{tab4}
\end{table*}
Fe K$_{\alpha}$ emission lines are detected in 6 (14) objects at a confidence
level larger than 2$\sigma$ (1$\sigma$). They
represent the first detection of such a feature in these objects. In
all cases, the centroid energies $E_l$ are consistent with emission
from neutral or mildly ionized iron. However, in UGC~1214 an
additional line with $E_l \simeq$6.66~keV suggests a contribution from
iron species as ionized as {\sc Fe xxv}. In Fig.~\ref{fig2}
the 5.25-7.25~keV spectra for the objects where a line is significantly
detected are shown. If the Fe K$_{\alpha}$ EW
is larger then and inconsistent with 600~eV
at the 90\% confidence level (the maximum
observed EW in ``Compton-thin'' objects; \cite{turner97,
risaliti02}) models assuming Compton-thick
obscuration have been preferred, whenever statistically
equivalent to those requiring a Compton-thin absorber.

Another powerful diagnostic of nuclear X-ray obscuration is
the 2--10~keV to O[{\sc iii}] flux ratio $T$.
Previous studies (\cite{maiolino98,bassani99,guainazzi05a})
have shown that objects with $T \le 0.1$ are invariably
Compton-thick, whereas objects with $T \ge 1$ are
almost exclusively
Compton-thin or unobscured.
For all objects where the iron line EW
gives a constraint on the nature of the nuclear
obscuration, this is consistent with that
inferred from the $T$ value. This gives us confidence in
using $T$ as a diagnostic of the nature of the X-ray
obscuration. The values of $T$ for the objects
of our sample are listed in
Tab.~\ref{tab3}.

A summary of the best-fit parameters and results is shown
in Tab.~\ref{tab2}, where the column density
\begin{table*}
\caption{Global fit continuum spectral results. ``$G_i$'' refers to the
Table number in this paper, where properties of the emission
lines detected in the global fits are
presented. The usage of two rows for a given objects
means that two thermal components are required by the fit.}
\begin{tabular}{lcccccc} \hline \hline
Source & $N_H$ & $f_s$ & $\Gamma$ & $kT$ & $G_i$ & $\chi^2/\nu$ \\
& ($10^{22}$~cm$^{-2}$) & (\%) & & (keV) & & \\ \hline
2MASSJ01500266-0725482 & $0.47 \pm^{0.05}_{0.04}$ & ... & $2.09 \pm^{0.13}_{0.07}$ & $0.10 \pm 0.02$  & 3 & 118.6/133 \\
ESO269-G012 & $9 \pm^9_5$ & ... & 2$^{\dag}$ & $0.76 \pm^{0.16}_{0.23}$ & & 0.8/1 \\
IC3639 & $>$160 & ... & $3.9 \pm^{2.5}_{1.1}$ & $0.37 \pm^{0.30}_{0.09}$ & 3 & 3.3/4 \\
IC3639 &  &  &  & $1.0 \pm^{8.4}_{0.8}$ &  &  \\
IC4995 & $>160$ & ... & $2.1 \pm0.9$ & $0.096 \pm^{0.013}_{0.014}$ & 3 & 15.4/11 \\
IRAS15480-0344 & $> 160$ & ... & 2$^{\dag}$ & $0.17 \pm^{0.07}_{0.09}$ & &  4.1/7 \\
Mkn1 & $>110$ & 1$^{\dag}$ & $2.41 \pm^{0.13}_{0.11}$ & ...  & 5 & 36.0/32 \\
Mkn266 & $>160$ & ... & 2$^{\dag}$ & $0.78 \pm^{0.11}_{0.15}$  & & 0.5/1 \\
Mkn612 & $65 \pm^8_6$ & $0.59 \pm^{0.26}_{0.15}$ & $2.40 \pm^{0.10}_{0.15}$ & $0.089 \pm^{0.017}_{0.008}$ & 3 &  32.6/23 \\
&  &  & & $0.62 \pm^{0.08}_{0.09}$ & & \\
NGC1410 & $>160$ & ... & $4.3 \pm^{0.6}_{0.5}$ & $0.61 \pm^{0.28}_{0.16}$ & & 7.2/6 \\
NGC17 & $40 \pm^{40}_{30}$ & $5 \pm^2_3$ & $1.85 \pm^{0.20}_{0.18}$ & $0.63 \pm^{0.09}_{0.11}$ & & 46.5/40 \\
NGC591 & $>160$ & ... & $2.4 \pm 0.9$ & $0.09 \pm 0.04$ & 3 & 29.6/29 \\
&  &  & & $0.69 \pm 0.05$ & & \\
NGC3982 & $>160$ & ... & 2$^{\dag}$ & $0.59 \pm^{0.13}_{0.18}$ & & 0.7/1 \\
NGC4156 & $<0.02$ & ... & $1.96 \pm^{0.08}_{0.05}$ & ... & & 207.3/179 \\
NGC4968 & $>160$ & ... & $2.4 \pm^{0.3}_{0.4}$ & ... & 3 & 23.7/23 \\
NGC5283 & $15 \pm 3$ & $0.9 \pm 0.5$ & $2.3 \pm 0.5$ & ... & & 12.5/13 \\
NGC5953 & ... & ... & $3.9 \pm^{1.4}_{1.3}$ & ... & & 0.6/1 \\
NGC7212 & $>160$ & ... & $1.5 \pm^{0.3}_{0.6}$ & $0.16 \pm^{0.15}_{0.09}$  & 3 & 97.0/65 \\
&  &  & & $0.72 \pm^{0.08}_{0.16}$ & & \\
NGC7465 & $46 \pm^{31}_{19}$ & $1.6 \pm^{0.9}_{1.3}$ & 2$^{\dag}$ & ... & & 18.0/24 \\
& $2.0 \pm^{0.6}_{0.5}$ & $9 \pm 3$ & & & \\
UGC1214 & $>160$ & ... & $2.7 \pm 0.4$ & $0.16 \pm 0.03$$^{\ddag}$ & 3 & 85.6/72 \\
&  &  & & $0.4 \pm^{0.4}_{0.2}$$^{\ddag}$ & & \\
UGC2456 & $90 \pm^{60}_{30}$ & ... & $1.97 \pm^{0.10}_{0.05}$ & ... & 3,5 & 60.7/48 \\
UGC2608 & $>160$ & ... & $2.3 \pm^{1.1}_{0.3}$ & $0.61 \pm^{0.14}_{0.46}$ & & 8.4/10 \\
& &  & & $1.5 \pm^{0.8}_{0.3}$ & &  \\
UGC4229 & $>160$ & ... & 2$^{\dag}$ & $0.59 \pm^{0.09}_{0.21}$  & & 9.4/7 \\
UGC6527 & $76 \pm^{21}_{17}$ & $0.7 \pm^{0.4}_{0.6}$ & $2.3 \pm 0.3$ & $0.71 \pm ^{0.14}_{0.12}$ & 3 & 15.3/22 \\
UGC987 & $0.068 \pm 0.013$ & ... & $1.70 \pm 0.04$ & ... & 3 &271.8/238 \\
UM625 & $>160$ & ... & $4.1 \pm 0.4$ & $0.23 \pm^{0.06}_{0.04}$ & & 8.1/6 \\
\hline \hline
\end{tabular}

\noindent
$^{\dag}$fixed

\noindent
$^{\ddag}$$Z = 0.19 \pm^{0.06}_{0.03}$

\label{tab2}
\end{table*}
for ``Compton-thick'' objects is
lower-bounded by $\sigma_t^{-1} \simeq 1.6 \times 10^{24}$~cm$^{-2}$.
The $EW$, $E_l$ and intensity $I_l$ of the Fe $K_{\alpha}$
fluorescent lines derived from the global fits are
shown in Tab.~\ref{tab4}, whereas the
same properties for other lines are listed
in Tab.~\ref{tab5} for the two objects (Mkn1 and
\begin{table}
\caption{$G_i$ not-iron emission lines detected  in
the sample. The EW are expressed again the scattering
continuum only.}
\begin{tabular}{lcc} \hline \hline
& Mkn1 & UGC2456 \\ \hline
\multicolumn{3}{l}{{\sc Fe-L}, 0.7--0.8~keV }\\
$E_l$ (keV) & $0.88 \pm 0.02$ & $0.805 \pm 0.016$ \\
$EW$ (keV) & $160 \pm^{60}_{40}$ & $140 \pm 40$ \\
$I_l$$^a$ & 2.9 & 1.1 \\
\multicolumn{3}{l}{{\sc Ne ix}, 0.915~keV} \\
$E_l$ (keV) & ... & $0.92 \pm 0.08$ \\
$EW$ (keV) & ... & $230 \pm^{50}_{30}$ \\
$I_l$$^a$ & ... & 1.4 \\
\multicolumn{3}{l}{{\sc Ne x}, 1.02~keV} \\
$E_l$ (keV) & ... & $1.026 \pm^{0.014}_{0.009}$ \\
$EW$ (keV) & ... & $180 \pm 40$ \\
$I_l$$^a$ & ... & 0.9 \\
\multicolumn{3}{l}{{\sc Fe-L}, 1.03--1.15~keV} \\
$E_l$ (keV) & $1.02 \pm 0.02$ & $1.15 \pm 0.02$ \\
$EW$ (keV) & $130 \pm^{30}_{50}$ & $100 \pm^{30}_{40}$ \\
$I_l$$^a$ & 0.6 & 0.6 \\
\multicolumn{3}{l}{{\sc Mg xi}, 1.34~keV} \\
$E_l$ (keV) & ... & $1.32 \pm 0.03$ \\
$EW$ (keV) & ... & $160 \pm^{20}_{70}$ \\
$I_l$$^a$ & ... & 0.46 \\
\multicolumn{3}{l}{{\sc S xiv}, 2.00~keV} \\
$E_l$ (keV) & ... & $1.90 \pm 0.06$ \\
$EW$ (keV) & ... & $120 \pm^{20}_{90}$ \\
$I_l$$^a$ & ... & 0.18 \\
\hline \hline
\end{tabular}

\noindent
$^a$in units of $10^{-5}$~counts~erg~s$^{-1}$

\noindent
$^b$fixed

\label{tab5}
\end{table}
UGC2456) where they are statistically requested.
 We stress that the meaning of
the EWs corresponding to the ``local'' and ``global'' fits 
in Tab.~\ref{tab4}
is not strictly the same. In the ``local'' fits, the $EW$ is
calculated against the background-subtracted
continuum underlying the line profile; in the ``global''
fits, it represent the EW against the proper continuum
to which it refers, {\it i.e.} the absorption corrected
primary power-law for Compton-thin, and the
Compton-reflection continuum for Compton-thick
objects.

\subsection{Notes on individual objects}

\noindent
{\it IC3639}, {\it NGC~1410}, and {\it UM~625}: 
in these objects $T < 0.1$, suggesting Compton-thick
obscuration. The EW of the K$_{\alpha}$ iron
line is basically unconstrained. However, the spectrum above
2~keV is typical of unobscured AGN, showing an {\it
observed} spectral index $\Gamma \simeq 2$. The hard X-ray
luminosity corresponding to the observed fluxes in these
sources is $\simeq$5--7$\times 10^{40}$~erg~s$^{-1}$. Longer
observations would be required to ultimately elucidate
the nature of the high-energy emission in these objects.
They can be reconciled with Compton-thick obscuration if
scattering is dominated by a hot plasma, or substantially
contaminated by serendipitous sources in the host 
galaxies, such as Ultra-Luminous
X-Ray binaries or bright Supernova Remnants.

\noindent
{\it NGC~7465}:
the application of the baseline model yields
a very flat intrinsic spectral index: $\Gamma \simeq 0.7$,
again unusual in AGN. A good fit is obtained if
one assumes that a ``standard'' ($\Gamma \equiv 2$) AGN
emission is covered by a partial covering absorber,
with column densities $N_{H,1} \simeq 5 \times 10^{23}$~cm$^{-2}$,
and $N_{H,2} \simeq 2 \times 10^{22}$~cm$^{-2}$, with
covering fractions of $15 \pm^{10}_{13} \%$ and
$85 \pm^{13}_{10} \%$, respectively.

\section{Discussion}

\begin{figure}
 \includegraphics[width=8cm]{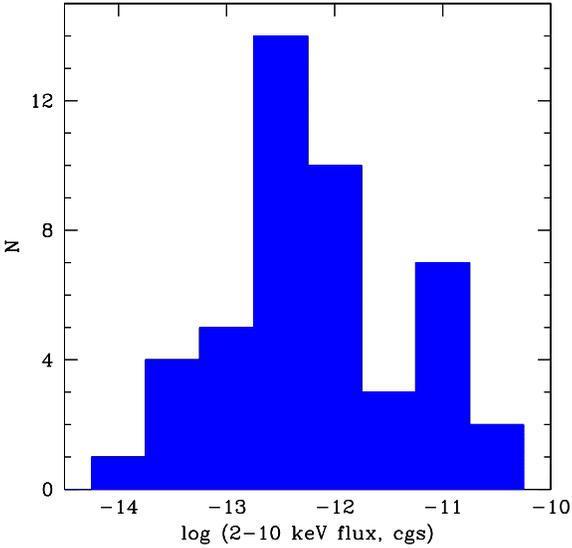}
\caption{2--10 keV flux for the objects or our sample}
\label{fig9}
\end{figure}

In this Section we combine the objects for which
new {\it Chandra}/XMM-Newton observations have been
presented in the previous Section, with measurements
extracted from the literature on all the other objects
of our sample. The sample covers a wide range of 2--10~keV
fluxes, from
$3 \times 10^{-11}$ to
$6 \times 10^{-15}$~erg~cm$^{-2}$~s$^{-1}$
(Fig.~\ref{fig9}).

\subsection{Correlation between X-ray absorption and
nuclear dust content}

\begin{figure}
 \includegraphics[width=8cm]{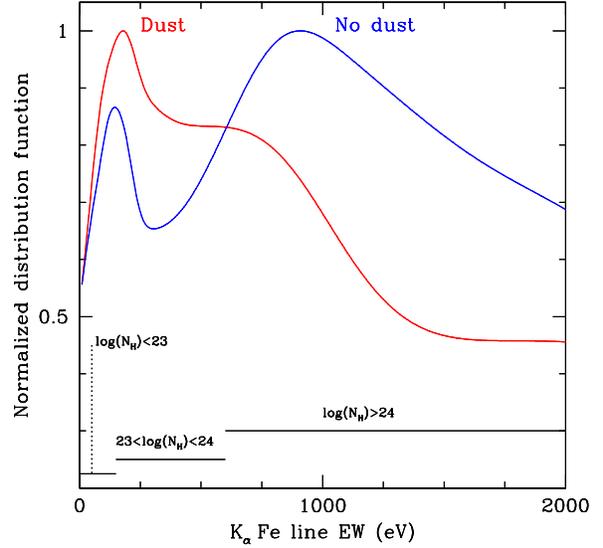}
\caption{Normalized distribution functions for
the Fe K$_{\alpha}$ line EW in ``dusty''
and ``not dusty'' Seyfert~2s. The {\it solid
bars} indicate typical EW values for
Compton-thin ($N_H \le 10^{23}$~cm$^{-2}$),
Extreme Compton-thin ($10^{23} < N_H < 10^{24}$~cm$^{-2}$),
and Compton-thick ($N_H \ge 10^{24}$~cm$^{-2}$)
objects (cf. Fig.~\ref{fig12})}
\label{fig10}
\end{figure}

In Fig.~\ref{fig10} we show normalized
robust distribution functions\footnote{They are histograms
of the distribution functions, where each measurement
is represented by a Gaussian function, whose
mean is the measurement value and whose deviation is the
measurement's statistical uncertainty} for the 
EW of the K$_{\alpha}$ iron line - used here
as an indicator
of X-ray obscuration -
for ``dusty'' and ``non-dusty''
Seyfert~2s. The probability that these
distributions are drawn from the same
parent population is $\sim 2 \times 10^{-8}$.
The distribution weighted mean
is $210 \pm 80$~eV, and $1000 \pm 300$~eV for
``dusty'' and ``non dusty'' galaxies, respectively.
The latter value is well
above the threshold separating Compton-thin from
-thick objects. This points to an average
systematic difference
in the X-ray absorbing column density depending
on the nuclear dust content in the host galaxy.

This difference is mainly due to objects with
a column density $10^{23}  \le N_H \le 10^{24}$~cm$^{-2}$
(``Extreme Compton-ThiN'', ECTN, objects hereafter).
The ratio $R$ between ECTN with and without
nuclear dust is $R_{ECTN} = 1.8 \pm 0.3$.
For ``Compton-ThicK'' (CTK) objects
the same ratio is a factor of
three lower (2$\sigma$ level): $R_{CTK} = 0.5 \pm 0.2$.
The ratio calculated on the whole sample is
0.77.

M98 define a further group of galaxies with interesting
dust absorption, as those highly inclined spirals with extensive
dust lanes on one side of their major axis and
hardly on the other side (``D-{\it [direction]''} in M98).
7 objects in our sample fall in this class.
Although they are not  ``nuclear dusty'' according
to our classification, it is interesting to observe that 
the correlation between X-ray obscuration and the presence of
dust becomes even more extreme if
we include them in the ``dusty'' sample:
$R_{ECTN} = 4.5 \pm 0.9$, $R_{CTK} = 1.0\pm 0.4$.
The ratio calculated on the whole sample is 1.56.

The existence of a correlation between
X-ray obscuration and dust
lanes in the HST images, points to a relation
between the host galaxy gas/dust, and the matter responsible
for X-ray obscuration in the innermost $\approxlt 100$~pc.
As already pointed out by G01, the correlation is probably not
direct, {\it i.e.} the dust lanes seen in high-resolution
HST images are not {\it directly} responsible for the
X-ray obscuration, as in ``DI'' objects these lanes do not
appear to cross the
line-of-sight to the active nucleus.
The connection must therefore be of an ``ambient'' or 
``evolutionary'' nature, with dusty nuclear environments
statistically favoring the formation of X-ray obscuring
patches and filaments.

The most straightforward interpretation of these
results is that Compton-thin obscuration,
primarily in the column density range
$10^{23} \le N_H \le 10^{24}$~cm$^{-2}$, is
due to gas associated to the dust lanes.
In other words: we measure a Compton-thick X-ray obscuration
when our line-of-sight intercepts
pc-scale compact clouds. We will still 
collectively dub them ``torus'' in the following,
although their detailed
geometry is largely unknown. Whenever
this does not occur, high-energy
photons have still a large likelihood to cross
gas associated with the dust
lane clouds, encompassing the nuclear environment
on a larger scale and with a larger covering fraction.

\begin{figure}
\hbox{
\hspace{0.25cm}
 \includegraphics[width=8.5cm]{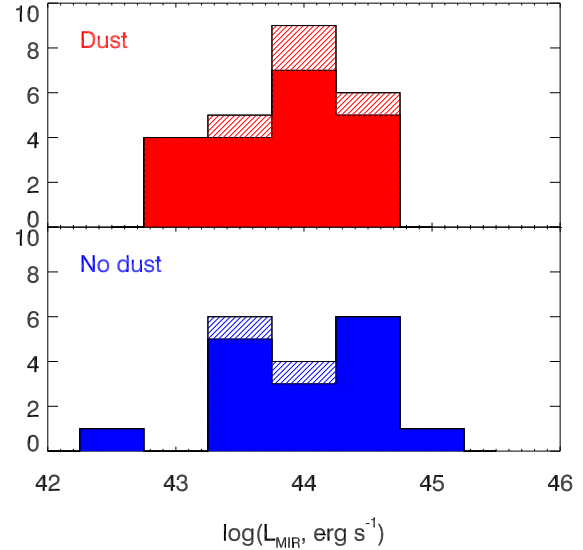}
}
\caption{MIR luminosity distribution histogram for
``dusty'' ({\it upper panel}) and ``non-dusty''
({\it lower panel}) objects in our sample.
{\it Shaded areas} indicate upper limits.}
\label{fig13}
\end{figure}

Alternatively, dust lanes and Compton-thick
``torii'' could be mutually exclusive in an AGN.
In this case, one might expect
that the distribution of the
Mid InfraRed (MIR) luminosity
differs between ``dusty'' and ``non
dusty'' Seyferts.
The MIR Spectral Energy Distribution of obscured
AGN is in facts dominated by ``torus'' reprocessing of the high-energy
continuum (\cite{alonsoherrero01,efstathiou05}).
In Fig.~\ref{fig13} we show such
distributions. The MIR flux density was calculated
from a simple linear combination of
12$\mu$ and 25$\mu$ IRAS measurements:
$$
S_{MIR} \equiv \nu_{12} S_{12} + \nu_{12} S_{25}
$$
The distributions are statistically indistinguishable.
A pc-scale Compton-thick ``torus'' is therefore most likely present
in all radio-quiet AGN, in agreement with the
ubiquity of unresolved iron fluorescent K$_{\alpha}$
emission lines in their X-ray spectra
(\cite{page03,jimenezbailon05}).

The viability of dust lanes as tracer of
Compton-thin X-ray obscuring gas 
depends on whether
galactic dust outside the nucleus can produce 
enough extinction. Estimates from HST color
maps (\cite{mulchaey94,simpson97,ferruit00,quillen99a})
suggest $A_V \simeq$0.5--1.5~mag. This corresponds
to $N_H \simeq$0.5--1.0$\times 10^{23}$~cm$^{-2}$ for gas-to-dust
ratios typically observed in luminous
and obscured nearby AGN
(\cite{granato97,maiolino01})\footnote{Interestingly,
the average surface density of molecular gas in the
innermost 500~pc of the Milky Way corresponds
to $N_H \simeq 4 \times 10^{22}$~cm$^{-2}$
(\cite{sanders84})}.
The fact that a large fraction of ``dusty'' Seyfert~2
in M98 have a rather disturbed morphology is suggestive
of clumpiness of the absorbing medium, which
may substantially increase the line-of-sight column
density in individual objects.

Maiolino et al. (1995) suggested the presence of
a ``100~pc torus'' in intermediate
({\it i.e.} 1.8--1.9) type Seyferts
to explain the fact they they are
preferentially found in edge-on hosts. Following
the results of their HST survey, M98 elaborated the ``Galactic
Dust Model'' (GDM) for obscuration - as opposed
to the standard ``Torus Model' -  where
the whole obscuration in type~2 AGN is explained
by the 100~pc dust lanes imaged by HST. However,
M98 themselves realized that this model
is not applicable to Compton-thick objects,
as dynamical considerations prevents the host
galaxy dust lanes from providing such extreme
obscuration. The GDM is therefore insufficient to
explain the large fraction of local obscured
AGN (\cite{risaliti99}; this paper). This motivated
Matt (2000) to propose an extension of the standard
unified scenario, whereby the
compact, dusty ``torus'' intercepts our line-of-sight
to Compton-thick AGN only, whereas Compton-thin
obscuration could be due to gas and dust located
at larger scales, possible related to the host
galaxy rather than to the nuclear engine.
\cite{guainazzi05a} and 
\cite{matt03b} have shown that
some Compton-thick objects would appear to us
as Compton-thin, if we could observe them through
a different line-of-sight. This supports the 
presence of a multiplicity of obscuring systems
in the innermost hundreds parsecs.

It must be emphasized that the existence of extended
obscuring structures on $\sim$100~pc scale as such
does {\it not} rule out the existence of an
obscuring torus. Compact toroidal geometries
(\cite{pier93}) are unable to explain the
broadness of the spectrum in the 1--20$\mu$m region,
and the nature of the 9.7$\mu$m silicate emission.
However, several different combinations of
geometry and torus physical properties are
consistent with the observed IR SED of highly
obscured AGN (\cite{galliano03}),
once allowance is made for
``extended torii'' (\cite{granato94}), where
the ratio between the outer and the inner radius
is $\simeq$10--100. Smooth (\cite{granato97})
or clumpy (\cite{nenkova02}) density distributions,
anisotropic flares  and tapered disk geometries
(\cite{efstathiou95}) are all equally successful in
reproducing the data. Moreover,
dust clouds might be associated
as well with the extended ionizing cones.
Mid-infrared
emission is slightly extended in many Seyfert~2
(\cite{maiolino95a,galliano03}).
Dust lanes
are sometimes associated with O[{\sc iii}] ionization
cones as well (\cite{quillen99a}).
Models including an optically thick torus
and an optically thin ``cone'' of dust clouds
(\cite{efstathiou95}) can as well reproduce the observed
IR SEDs (\cite{alonsoherrero03}).
Finally, radiation pressure driven
starburst disks could contribute as well to the
X-ray obscuration on scales $\approxlt 100$~pc
(\cite{thompson05}). All these possibilities can hardly
be discriminated even with the highest resolution
imaging detectors currently available, but could be
disentangled spectroscopically (\cite{lutz04}, and references therein).

\subsection{Overall sample properties}

The parent sample, from which
our sample discussed is extracted,
was selected
on the basis of observational properties, which should
not introduce any X-ray selection effects.
Our XMM-Newton program, from which most of the
XMM-Newton observations discussed in this paper are
drawn, carefully 
avoided any possible X-ray related bias. However,
the inclusion of archival {\it Chandra} and
XMM-Newton observations, belonging to different projects,
as well
as of the G01 objects, undoubtedly
may introduce a potential bias toward X-ray brighter,
and therefore potentially less obscured, objects.
With this caveat in mind, we review in the following
Section the main X-ray spectral properties of our
sample. 

\subsubsection{Soft excess}

\begin{figure}
\hbox{
\hspace{-0.25cm}
 \includegraphics[width=8.5cm]{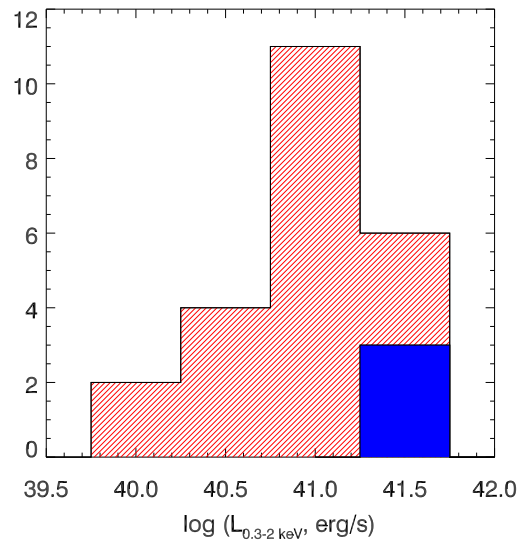}
}
\caption{Distribution function of the
0.3--2~keV intrinsic luminosity for the
sub-sample of our sample, observed by
either {\it
Chandra} or XMM-Newton. The {\it filled area}
correspond to objects, whose hard X-ray
emission is covered by a column density
$N_H \le 1.5 \times 10^{22}$~cm$^{-2}$,
implying contamination by the transmitted
nuclear emission in the soft X-ray band.}
\label{fig14}
\end{figure}

\begin{figure*}
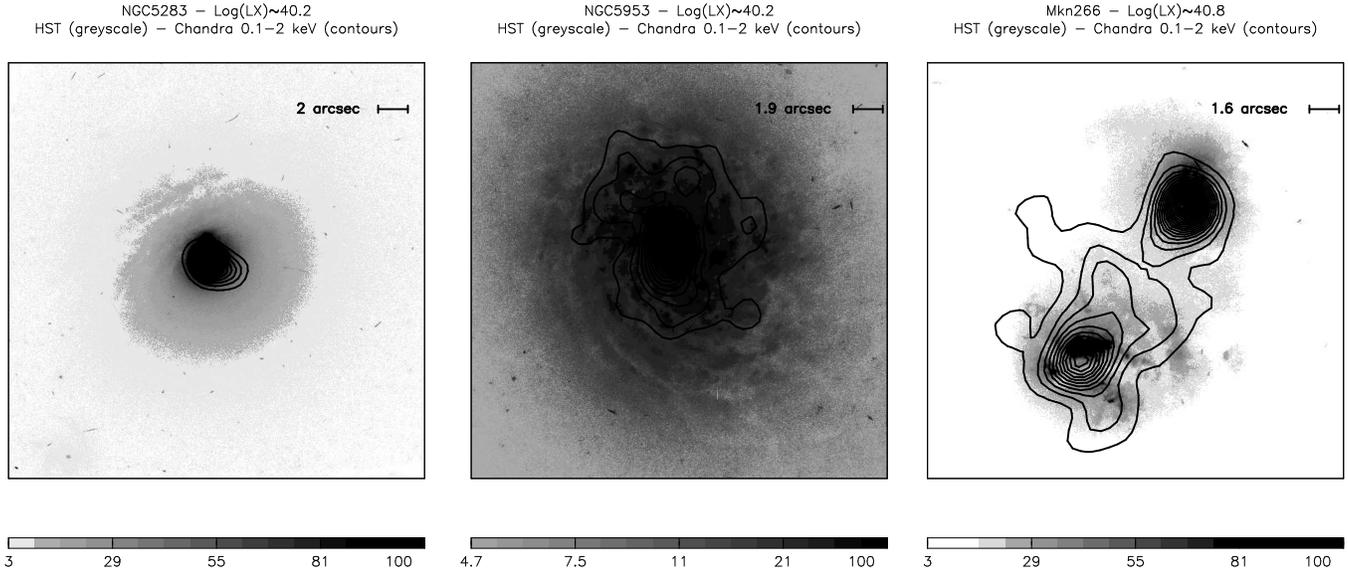

\hbox{
 \includegraphics[width=7.5cm,angle=-90]{fig15a.ps}
\hspace{0.25cm}
 \includegraphics[width=7.5cm,angle=-90]{fig15b.ps}
\hspace{0.25cm}
 \includegraphics[width=7.5cm,angle=-90]{fig15c.ps}
}
\caption{HST FW606 filter
({\it greyscale}) and {\it Chandra} 0.1--2~keV
images ({\it contours}) for: NGC~5283 ({\it left}),
NGC~5953 ({\it center}), Mkn~266 ({\it right}).
{\it Chandra} images were smoothed with a
0.75$\arcsec$ Gaussian kernel. Contours levels
are linearly spaced from 1 to the maximum of
measured pixel counts.}
\label{fig15}
\end{figure*}

The hard X-ray
continuum of type~2 Seyferts is dominated by a
power-law continuum, partly suppressed by photoelectric
absorption
(\cite{awaki91, turner97, risaliti02}).
In Compton-thick Seyferts, the transmitted continuum is
totally suppressed below 10~keV, and the hard X-ray continuum
is accounted for by reflection of the primary nuclear
emission off the inner walls of optically thick matter
surrounding the nuclear environment (\cite{matt00}).
The spectra of our sample -
including those objects whose X-ray spectra are presented
for the first time in this paper - are consistent with
this scenario.
Highly-obscured AGN exhibit almost invariably large soft
excess emission below 2~keV (\cite{guainazzi05a}).
In our sample, the 
total integrated soft X-ray luminosity of this
component range between $10^{40}$~erg~s$^{-1}$
and $4 \times 10^{41}$~erg~s$^{-1}$ (Fig.~\ref{fig14}).
In principle, this excess emission could be associated
with unresolved sources or diffuse emission in the
host galaxy, outshone by the formidable
output of the AGN in unobscured
objects. Although still tenable for some
individual objects, this explanation cannot
apply to the sample as a whole. {\it Chandra}
measurements of the luminosity function in
nearby spirals suggest that at most a few
sources with $L_X \sim 10^{40}$~erg~s$^{-1}$
are present in each galaxy (\cite{bauer01};
see \cite{wolter04} for an intriguing exception:
the Cartwheel galaxy). Likewise, there is
little - if any - diffuse emission associated with
spiral galaxies (at the level of a few $10^{39}$~erg~s$^{-1}$
in M~31; \cite{pietsch05,trinchieri04}).
We have searched for diffuse soft X-ray 
emission in
the 7 objects of our sample observed with
{\it Chandra}\footnote{The typical
Point Spread Function (PSF) of the
XMM-Newton observation correspond to $\ge$25~kpc radius}.
In 4 objects the radial profile is consistent
with the instrumental PSF.
In the two galaxies with the lowest 0.3--2~keV
luminosity (1.5--1.7$\times 10^{40}$~erg~s$^{-1}$)
we indeed detect a large-scale low-surface brightness
component ({\it left} ad {\it central panel} of
Fig.~\ref{fig15}), that in NGC~5953
extends across a significant fraction of the optical
surface of the host galaxy. Finally,
a peculiar object is Mkn~266
({\it right panel} of Fig.~\ref{fig15}), which belongs
to a close interacting galaxy pair, embedded in a luminous
(7$\times 10^{40}$~erg~s$^{-1}$) X-ray halo. A
detailed analysis of this intriguing
system is deferred to a future
paper.

In a few bright, nearby Seyfert~2s, for which
high resolution spectroscopic
measurements
have been possible (\cite{sako00, sambruna01, kinkhabwala02}),
soft X-rays are dominated by emission lines from highly
ionized species due mainly to recombination and
resonant scattering following photoionization and photoexcitation.
The density of the plasma is generally not large enough to
substantially contribute with a scattering continuum.
The striking coincidence between diffuse soft X-ray
and O[{\sc iii}] emission is consistent with this
scenario (Bianchi et al., in preparation).
However,
in some low-luminosity Seyfert~2s
($L_X \le 10^{40}$~erg~s$^{-1}$), high-resolution
imaging observations with {\it Chandra} have shown that
a contribution to the soft X-ray by plasma collisionally ionized
in episodes of intense nuclear star formation
is dominant in the soft X-ray regime (\cite{jimenezbailon03}).
It is interesting to observe that this soft component is generally
unobscured. It must be therefore produced in or by gas, extending
well beyond the material absorbing the nuclear continuum. This unfortunately
provides little constraint on its origin, as both nuclear
starbursts and AGN-driven photoionized extended structures on scales
of the order of several hundreds parsecs are common
in nearby AGN (\cite{wilson92, sako00, young01, bianchi03a}).

This ubiquitous soft excess may have an important impact
on the classification of Compton-thin/-thick objects at
cosmological redshift, if one uses standard X-ray survey colors.
This point is discussed in Appendix~A.

\subsubsection{K$_{\alpha}$ iron lines}

\begin{figure}
\hbox{
\hspace{-0.5cm}
 \includegraphics[width=9cm]{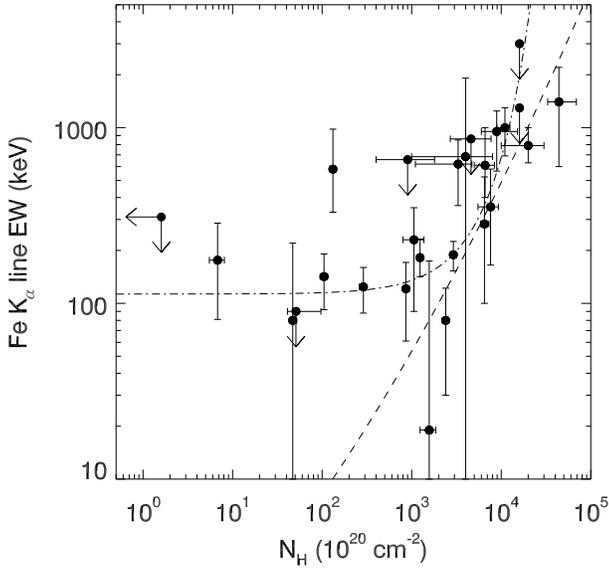}
}
\caption{X-ray obscuring column density against the
iron K$_{\alpha}$ fluorescent emission line EW
(details in text).
The {\it dashed line} shows the predicted EW from a uniform
shell of material encompassing the continuum source
(\cite{leahy93}). The {\it dot-dashed} line is
the best-fit in a scenario, where the
line is generated by reflection off the
inner wall of optically thick matter seen along
an unobscured line-of-sight, while the
underlying continuum is obscured by matter with a column
density $N_H$.}
\label{fig12}
\end{figure}

It has been early recognized that fluorescent K$_{\alpha}$
iron lines are a very common feature in the spectrum of
highly obscured AGN (\cite{awaki91}). In Fig.~\ref{fig12} we present
the EW of this feature (calculated against the
total continuum) versus the X-ray column densities
for all the objects of our sample, which do not
have a lower limit on their
column density, as in most cases this lower
limit was {\it inferred} on the basis of the high
iron line EW - as discussed in Sect.~3.
The two quantities are well correlated above
a column density $N_{H,tr} \sim 10^{23}$~cm$^{-2}$.
The correlation flattens to an almost constant
$EW_0$ for $N_H \le N_{H,tr}$.
The measured EWs are significantly
larger than expected
in transmission from a uniform shell of material
encompassing the continuum source (\cite{leahy93}).
On the other hand, obscuration of
continuum photons at $\simeq$6.4~keV
starts becoming significant at
column densities $\simeq N_{H,tr}$.
A censored data fit with a function:
$$
EW(N_H) = EW_0 e^{\sigma_{6.4} \times N_H}
$$
where $\sigma_{6.4}$ is the photoelectric cross-section
at 6.4~keV, yields an excellent fit: $\chi^2/\nu = 17.4/20$.
The function levels off at $EW_0 = 113 \pm 13$~keV.
This supports an origin of the iron line from 
the inner wall of the ``torus'' (\cite{ghisellini94, krolik94}).
It is interesting to observe that $EW_0$ is in good
agreement with the EW of the iron  K$_{\alpha}$
line observed in low-luminosity ($L_X \approxlt 10^{45}$~erg~s$^{-1}$),
X-ray unobscured quasars (\cite{jimenezbailon05}).
This suggests a common origin of this feature in
X-ray obscured and unobscured AGN (\cite{matt00}).

\subsubsection{$N_H$ distribution}

Compton-thick objects
\begin{figure}
 \includegraphics[width=8cm]{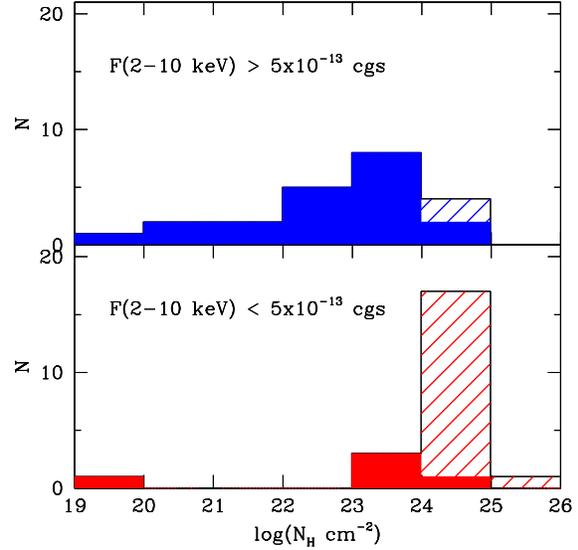}
\caption{$N_H$ distribution when our sample
is split according to the observed 2--10~keV flux}
\label{fig4}
\end{figure}
largely dominate the sample
below a 2--10~keV flux of
$5 \times 10^{-13}$~erg~cm$^{-2}$~s$^{-1}$.
There is, however, no
explicit dependency of $N_H$ on the intrinsic AGN luminosity.
If we use the O[{\sc iii}] flux as a proxy for the latter
quantity (it is impossible to have an estimate of the intrinsic
AGN luminosity from X-ray reflection-dominated spectra), no correlation
is observed with the measured column density
(Fig.~\ref{fig3}, see also \cite{risaliti99}),
\begin{figure}
 \includegraphics[width=8cm]{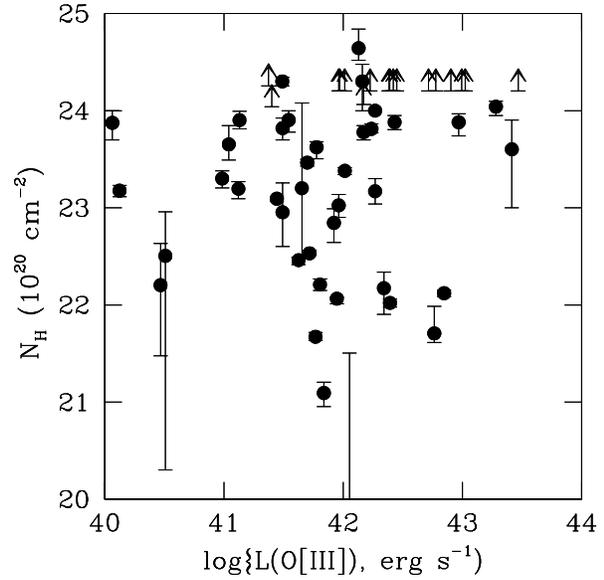}
\caption{O[{\sc iii}] luminosity versus column density
for the objects of our sample.}
\label{fig3}
\end{figure}
in agreement with the 0th order unified models.

The total fraction of Compton-thick
objects in our Seyfert~2 sample
is $46 \pm 10\%$. This is likely to represent a
{\it lower limit} to the fraction on Compton-thick
objects in the parent sample, due to our possible
bias toward X-ray brighter,
and therefore potentially less obscured, objects.

\section{Conclusions}

In this paper, we have presented new
low-resolution X-ray spectroscopic measurements for 26 nearby ($z < 0.035$)
Seyfert~2 galaxies. We join these results with information extracted from the
literature to build up a sample of 49 objects. The main goal of
this paper is the study of the correlation between the X-ray
column density and the 
dust content on $\approxlt 100$~pc scale as imaged by
the HST/WFC2. Our conclusions
can be summarized as follows.

\begin{itemize}

\item the presence of ``dust lanes'' is correlated with
X-ray obscuration
primarily in the column density range $10^{23} \le N_H \le
10^{24}$~cm$^{-2}$. We interpret this results as due to
the larger covering fraction of the gas associated with the
dust lanes, with respect to the compact, pc-scale,
Compton-thick ``torus''. Objects, whose line-of-sight
to the nucleus does not intercept the ``torus'',
still have a non-negligible chance to be absorbed
by gas on a larger scale
\\[0.25cm]
\item the fraction of Compton-thick objects
in our sample is 46$\pm$10\% and is strongly (observed-)
flux dependent, as expected. No
correlation exists between the amount of X-ray obscuration
and isotropic indicators of intrinsic AGN power in
the luminosity range covered by our sample
($L_X \approxlt 10^{43}$~erg~s$^{-1}$)
\\[0.25cm]
\item the spectrum of highly obscured AGN is complex,
with a prominent soft excess above the absorbed/reflected
nuclear continuum. This excess largely dominates the overall X-ray
{\it observed} energy output. This soft component
is generally unobscured. This suggests an origin
in gas structures, which extend beyond the material
responsible for the obscuration of the nuclear continuum.
The presence of this soft excess may confuse the estimate
of the column density in highly-obscured, $z \approxlt 1$
objects, when standard X-ray colors are used, as shown
in Appendix~A.

\end{itemize}

\begin{acknowledgements}
This paper is based on observations obtained with XMM-Newton, an ESA
science mission with instruments and contributions directly funded by
ESA Member States and the USA (NASA).
GM and GCP acknowledge financial support from MIUR under grant
{\sc cofin-03-02-23}.
This research has made use of
data obtained through the High Energy Astrophysics Science Archive
Research Center Online Service, provided by the NASA/Goddard Space
Flight Center and of the NASA/IPAC Extragalactic Database (NED) which
is operated by the Jet Propulsion Laboratory, California Institute of
Technology, under contract with the National Aeronautics and Space
Administration. We are grateful to M.Elitzur and
N.Loiseau for stimulating discussions, and to an
anonymous referee for valuable comments and suggestions.

\end{acknowledgements}

\section*{Appendix~A: Brief note on the usage of X-ray
colors to recognize Compton-thick objects}

\begin{figure*}
\hbox{
 \hspace{-0.5cm}
 \includegraphics[width=9cm, angle=90]{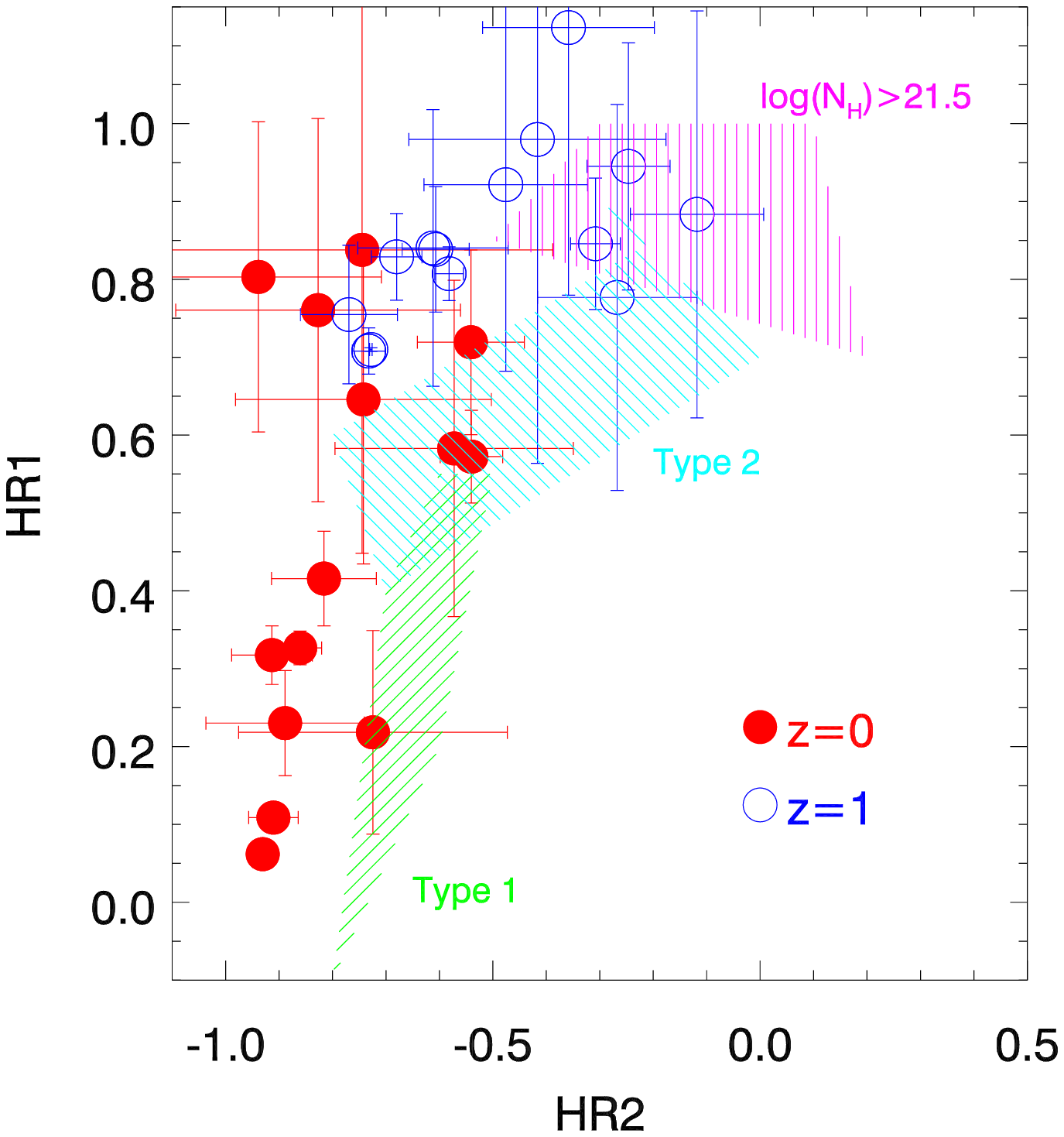}
 \includegraphics[width=9cm, angle=90]{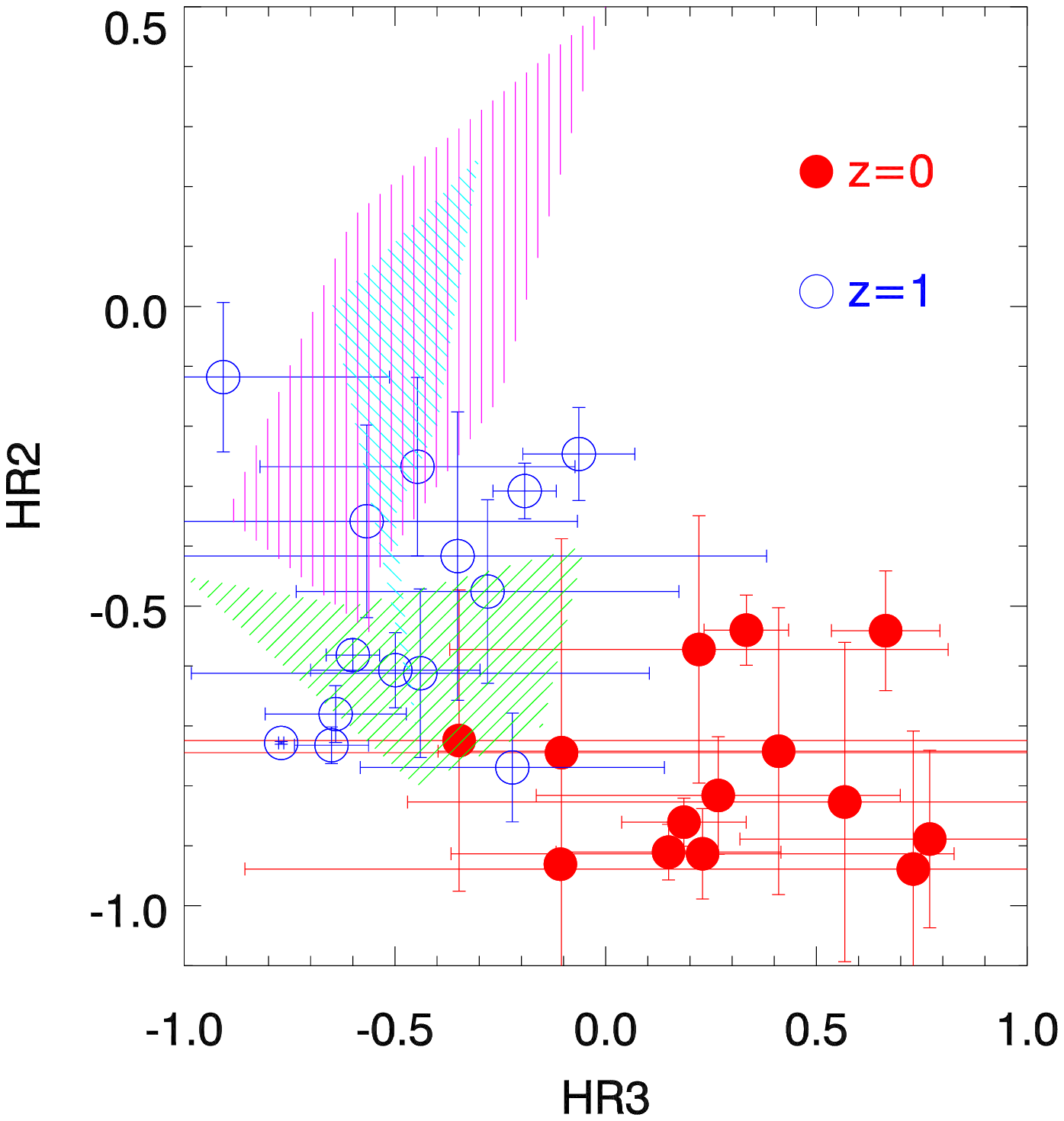}
}
\caption{Soft ({\it left panel}) and hard ({\it right panel})
color planes for the sample of Compton-thick discovered
in this paper, alongside the optically-defined 
Compton-thick sample of Guainazzi et al.
(2005a). The Hardness Ratios (HRs) are
define din text. The {\it shaded} areas represent
the loci occupied by type~1 ({\it left-bottom}
to {\it right-top shading}), type~2
({\it right-bottom}
to {\it left-top shading}), an
undefined highly obscured [$\log (N_H) > 10^{21.5}$]
sources ({\it vertical shading}; \cite{mainieri03}).
{\it Filled circles} correspond to Compton-thick
objects at their own redshift (``$z=0$'');
{\it empty circles} to the same objects, placed at
$z=1$}
\label{fig11}
\end{figure*}

Regardless of the detailed physical
origin of the soft excess, it is interesting to
estimate how such a component -
which can account for a significant fraction of the observed
flux in the X-ray band - may affect the identification of highly
obscured AGN at $z \approxlt 1$ (at larger redshifts this component
is shifted outside the energy bands where X-ray surveys
are conducted).
In Fig.~\ref{fig11} we show color plots for the sample of Compton-thick AGN
discovered by the analysis described in our sample (Sect.~4)
together with the optically-defined sample of Compton-thick
Seyfert~2 galaxies discussed by Guainazzi et al. (2005a;
the Circinus Galaxy has been excluded from the sample,
as its soft X-ray emission is substantially suppressed
by a large Milky Way's intervening column density).
For each
source, we have measured the count rates in 4 different
energy bands: 0.2--0.5~keV (Ultra-Soft; US), 0.5--2.0~keV
(Soft; S), 2.0-4.5~keV (Medium, M), and 4.5--10.0~keV
(Hard, M). We have
calculated the following Hardness Ratios (HRs: \cite{hasinger01}):
$HR1 \equiv (S-US)/(S+US)$, $HR2 \equiv (M-S)/(M+S)$, 
$HR3 \equiv (H-M)/(H+M)$. Fig.~\ref{fig11} shows
the {\it soft color plane} ($HR2$ versus $HR1$)
and the {\it hard color plane} ($HR3$ versus $HR2$)
for Compton-thick ``local'' galaxies, together with
the loci occupied by ``type~1'' AGN, ``type~2'' AGN
and unclassified highly obscured ($\log (N_H) > 10^{21.5}$)
sources
in the XMM-Newton observation of the Lockman Hole
(\cite{mainieri03}).
Compton-thick Seyfert~2 galaxies occupy a specific
area of these planes, thanks to their combination of
large $HR3$ (flat hard X-ray spectrum) and low $HR2$
(steep soft X-ray excess). Interestingly enough, they
span a range in $HR1$ which encompasses typical values for
both ``type~1'' and ``type~2'' AGN, suggesting a multiplicity
of possible origins for the soft excess emission. In the
same Figure we show which loci our ``local''
Compton-thick sources would occupy, if they were located at $z=1$.
In the hard color plane, Compton-thick Seyfert~2 galaxies
would be basically indistinguishable from other AGN types.
A sizable fraction of them would actually occupy the same
region as broad-line AGN. The soft color plane would
be required to remove the degeneracy, although
Compton-thick Seyfert~2s would
still have a softer $HR2$ with respect to other less obscured AGN,
which could lead to an underestimate of the column density covering
their active nucleus, and consequently of the
fraction of highly obscured AGN at $z \simeq 1$.

\end{document}